\documentclass[12pt,oneside]{article}
\usepackage[spanish]{babel}
\usepackage{amssymb,epsfig}
\usepackage{color}

\begin{document}

\pagecolor{white}

\vskip 60pt


\centerline{\LARGE\textbf{\textcolor{red}{\bf
Principio de invariancia de norma y}}}
\medskip
\centerline{\LARGE\textbf{\textcolor{red}{\bf rompimiento espont\'aneo de
simetr\'{\i}a}}}
\medskip
\centerline{\LARGE\textbf{\textcolor{red}{\bf en una part\'{\i}cula
cl\'asica}}}
\bigskip

\centerline{J. Mahecha G.${}^*$, L. A. S\'anchez D.${}^\dagger$${}^*$}
\centerline{\it ${}^*$Instituto de F\'{\i}sica, Universidad de
Antioquia, Medell\'{\i}n, Colombia\/}
\centerline{\it ${}^\dagger$Escuela de F\'{\i}sica,
Universidad Nacional de Colombia, Sede Medell\'{\i}n\/}
\centerline{Correo e: {\tt mahecha@fisica.udea.edu.co,
lasanche@perseus.unalmed.edu.co}}
\centerline{Marzo 2003}
\bigskip

 RESUMEN. Debido a que s\'olo los campos de materia tienen fase,
 con frecuencia se piensa que el principio de invariancia de norma
 puede inducir campos de norma solamente en sistemas
 cu\'anticos. Pero esto no es necesario. En este
 art\'{\i}culo, de car\'acter pedag\'ogico, se
 presenta un sistema cl\'asico, constituido por una
 part\'{\i}cula en un potencial, que luego se usa
 como mo\-de\-lo para ilustrar el principio de invariancia de norma y
 el rompimiento espont\'aneo de simetr\'{\i}a.
 Estos conceptos aparecen en el estudio de las
 transiciones de fase de segunda clase. Entre los
 fen\'omenos que presentan estas transiciones
 est\'an: La ferroelectricidad, el ferromagnetismo,
 la superconductividad, los plasmones en un gas de
 electrones libres, y la descripci\'on de la masa
 de los bosones vectoriales en las teor\'{\i}as
 de campos de norma de Yang-Mills.\\

 ABSTRACT. Due to the fact that only matter fields
 have phase, frequently is believed that the gauge principle
 can induce gauge fields only in quantum systems.
 But this is not necessary. This paper, of pedagogical
 scope, presents a classical system constituted by a
 particle in a classical potential, which is used as a
 model to illustrate the gauge principle and the
 spontaneous symmetry breaking. Those concepts appear
 in the study of second order phase transitions.
 Ferroelectricity, ferromagnetism, superconductivity,
 plasmons in a free electron gas, and the mass of
 vector bosons in the gauge field Yang-Mills theories,
 are some of the phenomena in which these transitions occur.\\


\newpage

\section{\textcolor{red}{Introducci\'on}}
\label{sec:intro}

El principio de invariancia de norma, de acuerdo al cual cada una de las
interacciones fundamentales entre fermiones elementales es mediada por el
intercambio de los bosones de norma del grupo de simetr\'{\i}a correspondiente,
est\'a en la base de las teor\'{\i}as modernas de part\'{\i}culas
elementales~\cite{salam,weinberg}. Su
importancia es tal que cualquier modelo que pretenda hacer una descripci\'on
realista (renormalizable) de las interacciones fundamentales, debe primero
asegurar que se cancelan todas las posibles anomal\'{\i}as que pueden poner en
peligro su validez~\cite{ward}.

El grupo de simetr\'{\i}a de una interacci\'on fundamental puede ser abeliano como
en el caso del  grupo $U(1)_Q$  de transformaciones de fase de la
electrodin\'amica; o puede contener factores no abelianos como en el caso
del  grupo $SU(3)_c  \otimes SU(2)_L  \otimes U(1)_Y$ del  modelo de
Wienberg-Salam  que describe  las interacciones  fuertes ($SU(3)_c$) y
electrod\'ebiles ($SU(2)_L \otimes U(1)_Y$).
Teor\'{\i}as basadas en grupos
no abelianos de transformaciones de norma locales son llamadas teor\'{\i}as de
Yang-Mills~\cite{yang}.
Si los par\'ametros asociados a un
grupo  de transformaciones  de  simetr\'{\i}a  son  independientes de  la
localizaci\'on espacio-temporal la simetr\'{\i}a es global; en caso contrario es
local. La simetr\'{\i}a, a su vez, puede ser exacta como la descrita por el grupo
$SU(3)_c$ del color, caso en el cual los bosones de norma correspondientes
son no masivos; o puede estar espont\'aneamente rota (oculta) debido a que
campos escalares, llamados campos de Higgs, adquieren valores esperados en
el vac\'{\i}o distintos de cero que hacen que el estado vac\'{\i}o de la subyacente
teor\'{\i}a cu\'antica de campos no sea invariante de norma aunque s\'{\i} lo sea el
Lagrangiano. El teorema de Goldstone~\cite{goldstone} asegura entonces la aparici\'on de campos
escalares sin masa asociados a grados de libertad no f\'{\i}sicos que ser\'an
absorbidos por los bosones de norma del grupo de simetr\'{\i}a rota, los cuales,
como consecuencia, adquieren masa. Este fen\'omeno es el llamado mecanismo de
Higgs~\cite{anderson}. En las referencias~\cite{moriyasu} se presentan
los detalles de la teor\'{\i}a general de las transformaciones de norma.

El presente trabajo tiene como finalidad ilustrar estas ideas con un sistema
cl\'asico muy simple en el cual se presenta invariancia bajo transformaciones
de norma abelianas.

Sea una esfera de masa $m$ que puede oscilar a lo largo del eje $y$, sujeta a un
par de resortes. Su energ\'{\i}a potencial es $U(y)=-Cy^2+B y^4$.
Adicionalmente se la somete a una fuerza oscilante de la forma
$f_0\cos\gamma t$, donde $\gamma$ es mucho mayor que la frecuencia de las
oscilaciones libres. La oscilaci\'on forzada tiene una amplitud $y_0$. Al
hallar la frecuencia de peque\~nas oscilaciones (lineales) en funci\'on
del par\'ametro $T=y_0^2$ se encuentra que hay un valor $T_C$ en el cual
esta frecuencia var\'{\i}a de manera discontinua \cite{serbo}.

Luego el modelo se modifica de la siguiente manera: Se toma una
part\'{\i}cula cargada en un potencial $U(\rho)$ similar a $U(y)$, con
simetr\'{\i}a cil\'{\i}ndrica, y tambi\'en sometida al efecto de la fuerza
oscilante de alta frecuencia, cuando $T<T_C$, y adem\'as se coloca un
campo magn\'etico uniforme. Finalmente se analizan los efectos
relacionados con la elecci\'on de la norma del potencial del campo
electromagn\'etico.

En este art\'{\i}culo se argumenta la presencia del ``mecanismo de
Higgs'', uno de los principales ingredientes del modelo st\'andard, en el
sistema cl\'asico descrito. Este trabajo tiene relaci\'on con otros
aparecidos recientemente en la literatura. Se ha observado el rompimiento
espont\'aneo de simetr\'{\i}a en mec\'anica cl\'asica de part\'{\i}culas
descrita por medio de una funci\'on de distribuci\'on \cite{ogawa},
igualmente en la simple colocaci\'on de un bloque rectangular sobre una
cu\~na formada por dos superficies \cite{acus}. El problema de la
invariancia de norma y su relaci\'on con las variables can\'onicas ha sido
tambi\'en estudiado \cite{khriplovich}. T. Dittrich {\it et al}
\cite{dittrich} han estudiado efectos debidos a la adici\'on de un
t\'ermino oscilante sobre el movimiento de part\'{\i}culas en un potencial
del tipo ``diente de sierra''.

\section{\textcolor{red}{Modelo mec\'anico}}
\label{sec:modelo}

\begin{figure}[h]
\vspace{-10mm}
\centerline{\psfig{figure=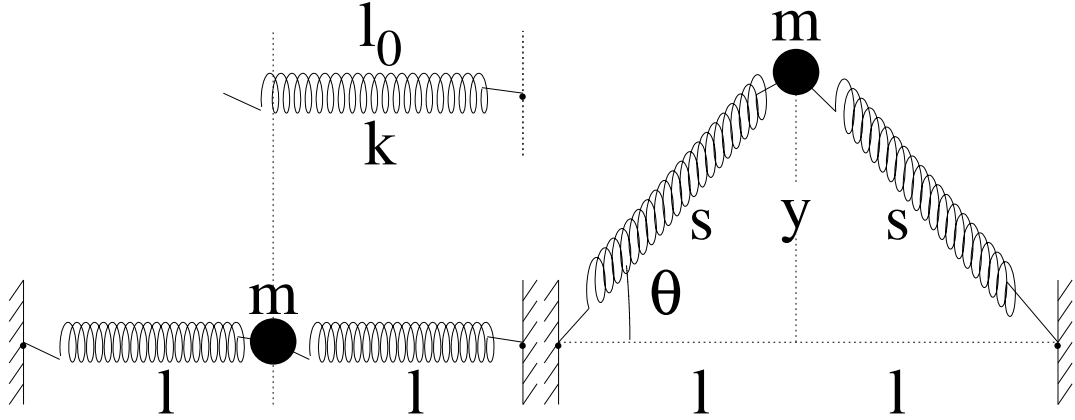,width=12cm,angle=270}}
\vspace{-30mm}
\caption[Short Title]{Sistema no lineal formado por una masa y dos
resortes.}
\label{fig:resortes}
\end{figure}

Se asume que $m$ y $k$ son tales que se puede ignorar el
efecto del peso de $m$, y que $l_0>l$ ($l_0=l+d$). V\'ease
Fig. \ref{fig:resortes}. La fuerza neta vertical sobre la
esfera es \cite{marion}
\begin{equation}
F=-2k(s-l_0)\sin\theta=-2k(s-l-d)\sin\theta.
\end{equation}

Se reemplaza $s=(l^2+y^2)^{1/2}$ y $\sin\theta=y/(l^2+y^2)^{1/2}$,
con lo cual
\begin{equation}
F=\frac{2kd}{l}y-k\frac{d+l}{l^3}y^3.
\end{equation}

Se ve que cuando $d=0$ el resorte es intr\'{\i}nsecamente no
lineal.

La energ\'{\i}a potencial est\'a dada por
\begin{equation}
\begin{array}{rcl}
U(y)&=&\displaystyle
-k\frac{l_0-l}{l}y^2+k\frac{l_0}{4l^3}y^4\\
&&\\
&=&\displaystyle 
-Cy^2+By^4.
\end{array}
\label{eq:erroneo}
\end{equation}
$C$ y $B$ son positivas. Podr\'{\i}amos pensar que al orden m\'as bajo,
$U(y)=-Cy^2$ representa un ``oscilador arm\'onico'' con masa de signo
err\'oneo.

Si $x_e$ es una posici\'on de equilibrio de $U$, se tiene
para $y\approx x_e$,
\begin{equation}
\begin{array}{rcl}
U(y)
&\approx&\displaystyle 
U(x_e)+\frac{1}{2}U''(x_e)(y-x_e)^2,\\
&&\\
\displaystyle -\frac{dU(y)}{dy}&\approx&-U''(x_e)(y-x_e).
\end{array}
\end{equation}

De $U'=-2Cy+4By^3=y(-2C+4By^2)$ y $U''=-2C+12By^2$ se sigue
que las posiciones de equilibrio posibles son
\begin{equation}
\begin{array}{rcl}
x_0&=&0\\ &&\\ &{\rm y}&\\ &&\\ x_{\pm}&=&\displaystyle
\pm\left(\frac{C}{2B}\right)^{1/2},
\end{array}
\end{equation}
en las cuales la segunda derivada vale $U''(x_0)=-2C<0$,
con lo cual $x_0$ es inestable, y $U''(x_\pm)=4C>0$, con lo
tanto $x_-$ y $x_+$ son estables.

\begin{figure}[h]
\vspace{0mm}
\centerline{\psfig{figure=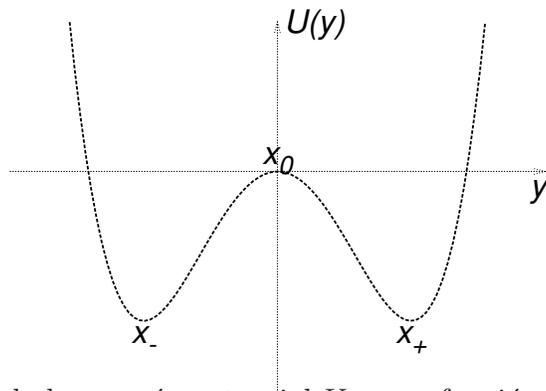,width=6cm,angle=270}}
\vspace{-10mm}
\caption[Short Title]{Dibujo de la energ\'{\i}a potencial $U$ como
funci\'on de $y$. Como $x_-=-x_+$, $U$ tiene la simetr\'{\i}a $y\to-y$,
pero la oscilaci\'on estable no posee esta simetr\'{\i}a.}
\label{fig:potencial}
\end{figure}

Por lo tanto, para movimientos cercanos a $y=x_0$ y $y=x_\pm$ se tiene
\begin{equation}
\displaystyle
 -\frac{dU(y)}{dy}\approx+2Cy,\
 -\frac{dU(y)}{dy}\approx-4C(y-x_\pm),
\end{equation}
respectivamente. El alejamiento indefinido de $x_0$ es caracterizado por
$\lambda=(2C/m)^{1/2}$ y las peque\~nas oscilaciones alrededor de $x_\pm$
por la frecuencia $(4C/m)^{1/2}$.

Ahora, se somete la esfera una fuerza
\begin{equation}
f(t)=f_0\cos\gamma t,
\end{equation}
la cual induce oscilaciones forzadas $-y_0\cos\gamma t$, donde
$f_0=m(\lambda^2+\gamma^2)y_0$. N\'otese que $U(y)$ no admite peque\~nas
oscilaciones libres alrededor de $x_0$ y que $y_0$ decrece mon\'otonamente
con $\gamma$. Esto difiere del caso en el cual la posici\'on de equilibrio
$x_0$ es estable, $y_0$ tiene un m\'aximo en la frecuencia de resonancia,
y hay un desfase de $\pi$ de la oscilaci\'on forzada respecto a la fuerza
externa. V\'ease \S 22 en \cite{landaumec}.

La ecuaci\'on del movimiento es
\begin{equation}
m\ddot y=-\frac{dU(y)}{dy}+f(t).
\end{equation}

\section{\textcolor{red}{``Transici\'on de fase''}}
\label{sec:phasetra}
Se asume ahora que $y=x+y_0\cos\gamma t$, donde $x$ es la componente de
bajas frecuencias del desplazamiento \cite{landaumec}.
\begin{equation}
m\ddot x -my_0\gamma^2\cos\gamma
t=-\frac{dU(y)}{dx}+f_0\cos\gamma t.
\label{eq:ecmov}
\end{equation}

Promediando los dos lados de la ecuaci\'on del movimiento
(\ref{eq:ecmov}) en un per\'{\i}odo de la oscilaci\'on de
alta frecuencia se obtiene
\begin{equation}
m\ddot x= -\left\langle\frac{dU(x+y_0\cos\gamma
t)}{dx}\right\rangle,
\end{equation}
donde
\begin{equation}
\displaystyle
\frac{dU}{dx}=2C(x+y_0\cos\gamma t)+4B(x+y_0\cos\gamma t)^3.
\end{equation}
Usando $\langle\cos\gamma t\rangle=\langle\cos^3\gamma
t\rangle=0$ y $\langle\cos^2\gamma t\rangle=1/2$ se llega a
\begin{equation}
\begin{array}{rcl}
m\ddot x&=&2Cx-4Bx^3-6By_0^2x\\ &=&2(C-3By_0^2)x-4Bx^3.
\end{array}
\end{equation}
La fuerza y la energ\'{\i}a potencial efectivas son
\cite{landaumec}
\begin{equation}
\begin{array}{rcl}
F_{ef}&=&2(C-3By_0^2)x-4Bx^3\\
U_{ef}&=&-(C-3By_0^2)x^2+Bx^4=Ax^2+Bx^4,
\end{array}
\end{equation}
donde $A$ puede ser positiva o negativa al cambiar $y_0$.
Adem\'as $U_{ef}=U$ para $y_0=0$.

Si se definen las cantidades positivas $T=y_0^2$ y
$T_C=C/(3B)$,
\begin{equation}
U_{ef}=-3B(T_C-T)x^2+Bx^4.
\end{equation}
$U_{ef}$ tiene la misma forma de $U$ con un diferente coeficiente de
$x^2$. La figura \ref{fig:potencial} sirve para representar a $U_{ef}$
para $T<T_C$. En el modelo de la figura \ref{fig:resortes}, el sistema
din\'amico obtenido de $U$ puede sufrir una {\it bifurcaci\'on\/} que da
lugar a la aparici\'on de dos puntos de equilibrio estables cuando el
origen se desestabiliza, cuando se var\'{\i}a el par\'ametro $d$. Pero
acabamos de encontrar otra forma de realizar lo mismo, variando $T$ o sea
la amplitud de la fuerza aplicada de frecuencia muy alta.

\begin{figure}[t]
\vspace{-10mm}
\psfig{figure=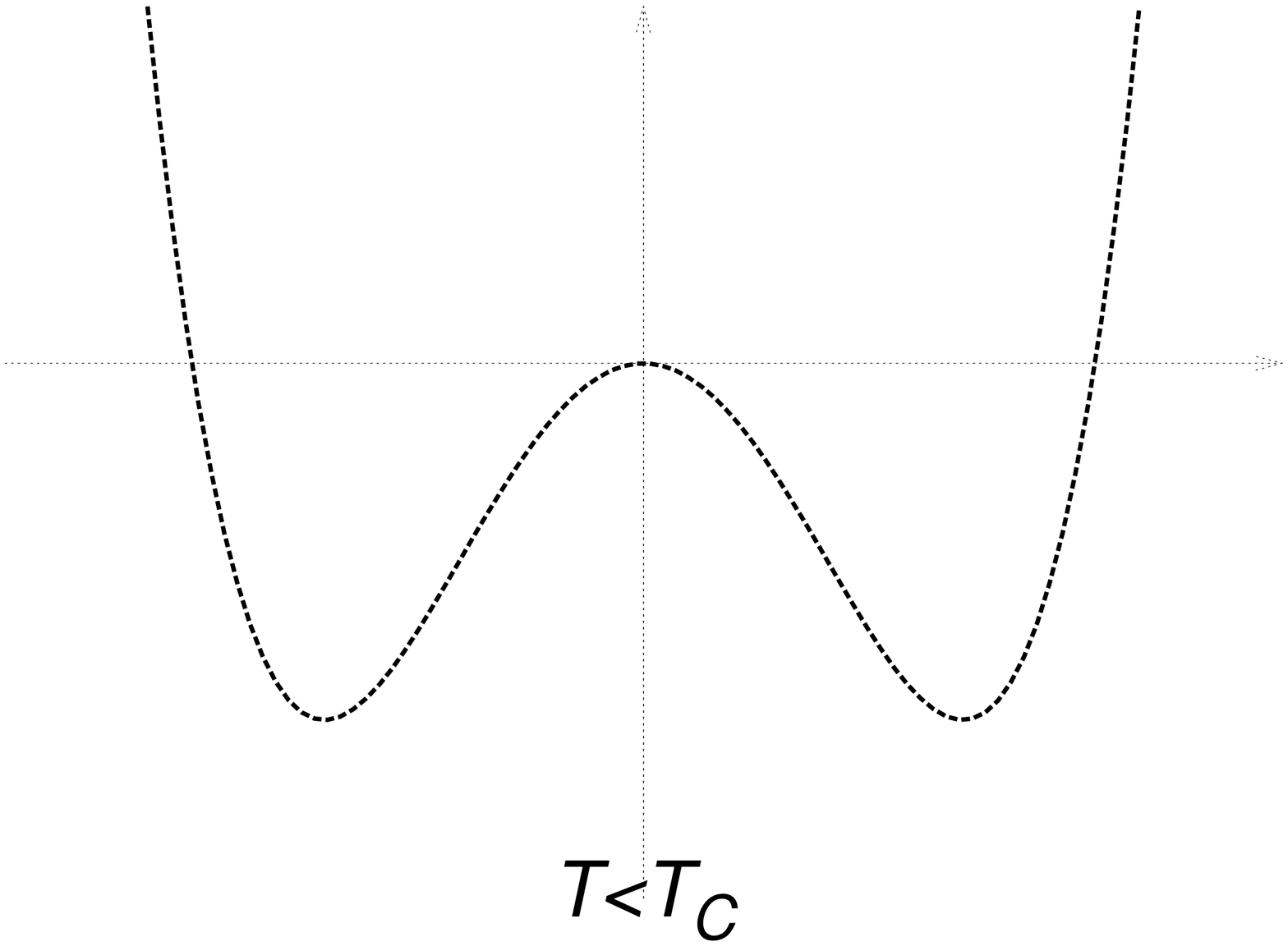,width=5cm,angle=270}
\hfill\psfig{figure=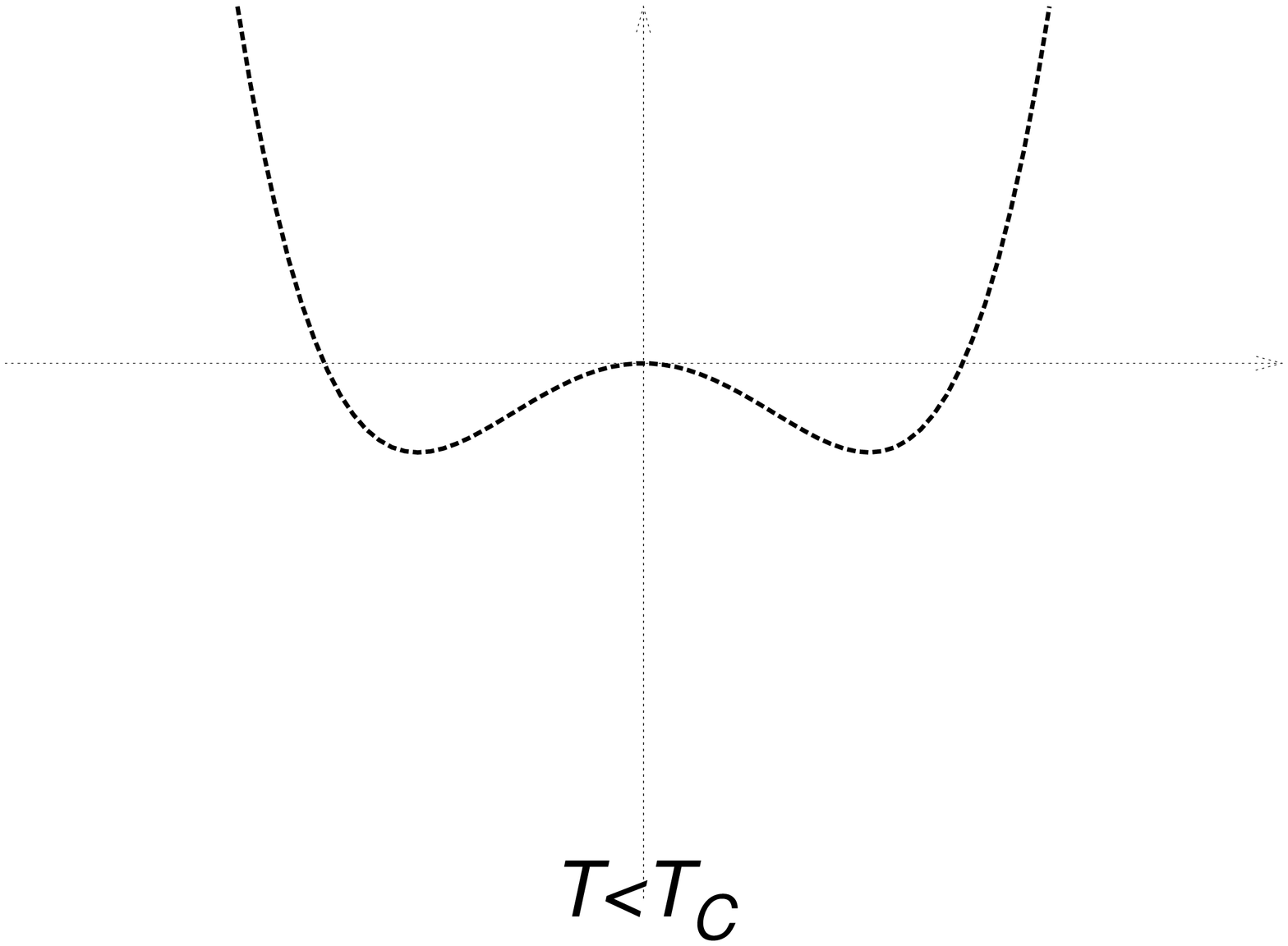,width=5cm,angle=270}
\vspace{10mm}
\psfig{figure=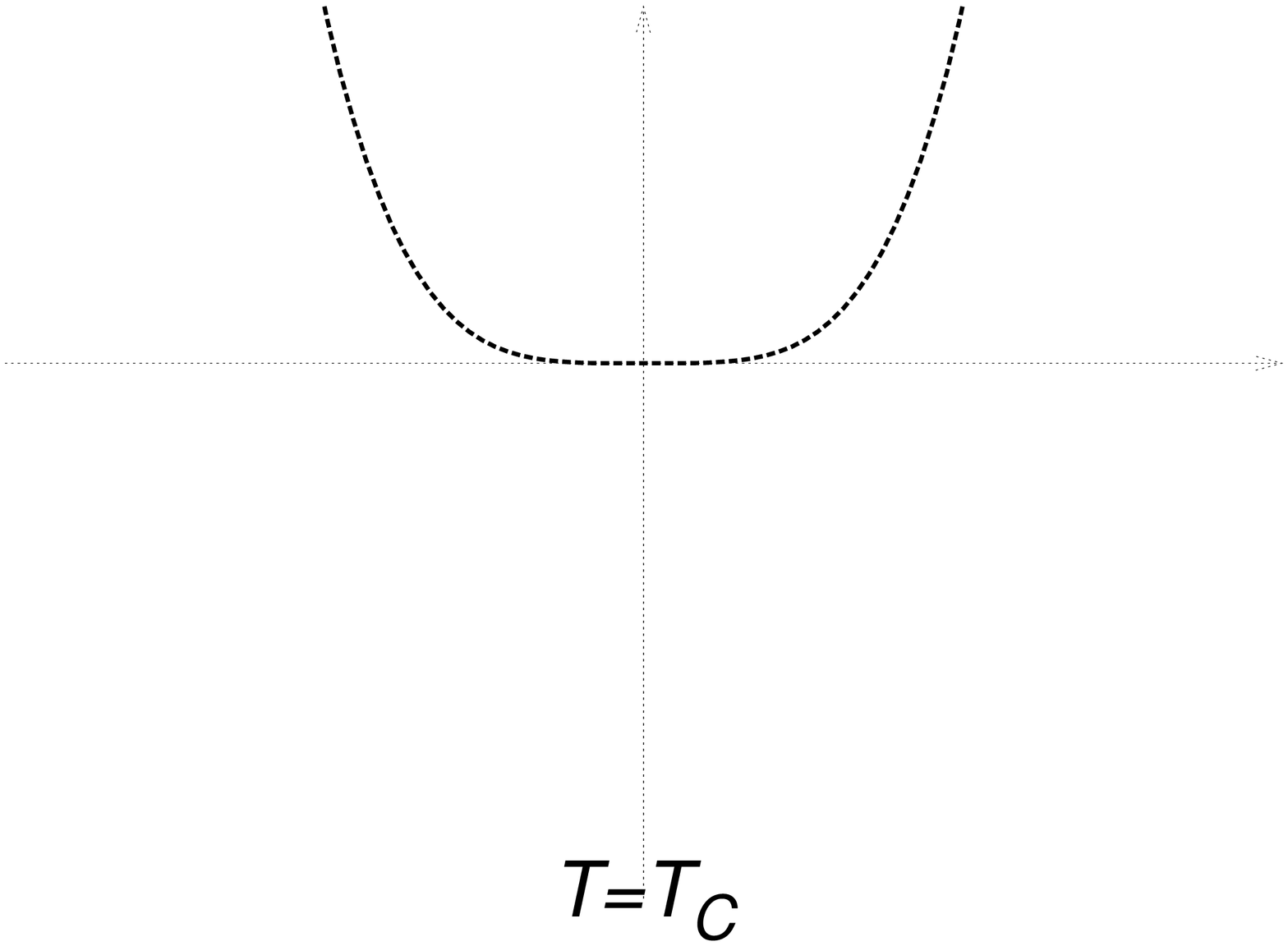,width=5cm,angle=270}
\hfill\psfig{figure=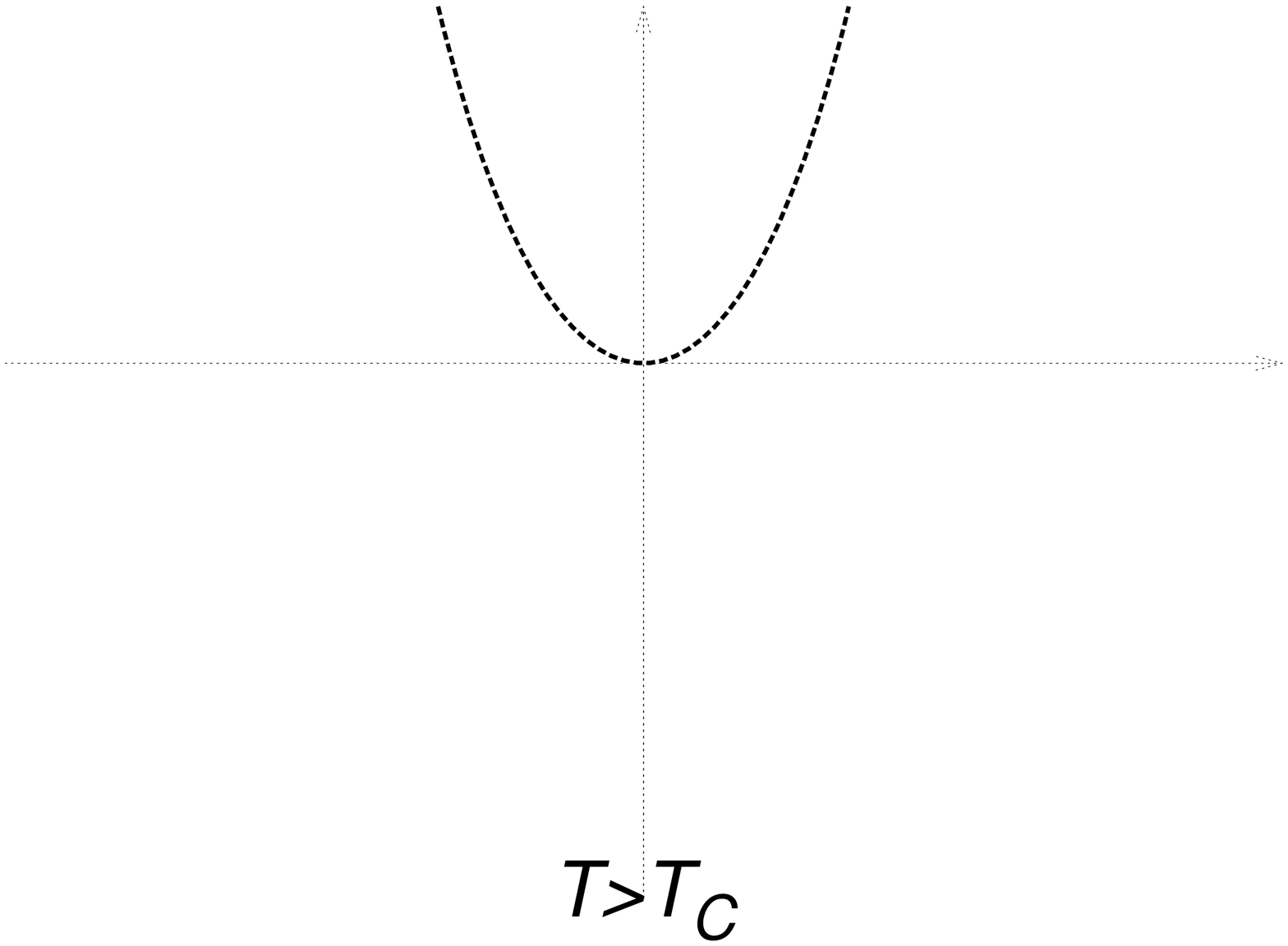,width=5cm,angle=270}
\vspace{-10mm}
\caption[Short Title]{Deformaci\'on de $U_{ef}$ cuando $T$
aumenta continuamente desde va\-lo\-res $T<T_C$ hasta
$T>T_C$. En $T=T_C$ el oscilador es intr\'{\i}nsecamente
no lineal.}
\label{fig:transicion}
\end{figure}

Cuando $T<T_C$ las posiciones de equilibrio $x_\pm$ son
estables y $x_0$ es inestable, pero cuando $T>T_C$ solo
existe una posici\'on de equilibrio, $x_0$, la cual es
estable. V\'ease la figura \ref{fig:transicion}.

Si $T<T_C$ la frecuencia de peque\~nas oscilaciones
alrededor de $x_\pm$ es
\begin{equation}
\omega'=\left(-2\frac{A}{m}\right)^{1/2}=(6B)^{1/2}(T_c-T)^{1/2},
\end{equation}
y cuando $T>T_C$ la frecuencia de peque\~nas oscilaciones
alrededor de $x_o$ es
\begin{equation}
\omega'=\left(2\frac{A}{m}\right)^{1/2}=(6B)^{1/2}(T-T_C)^{1/2}.
\end{equation}

\begin{figure}[t]
\vspace{10mm}
\centerline{\psfig{figure=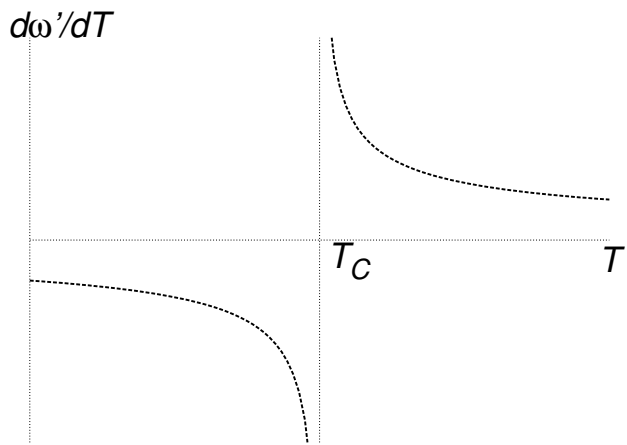,width=6.5cm,angle=270}}
\vspace{0mm}
\caption[Short Title]{Discontinuidad de la derivada de la
frecuencia en $T=T_C$.}
\label{fig:salto}
\end{figure}

La derivada de la frecuencia $\omega'$ respecto a $T$ tiene
un salto en $T=T_C$. V\'ease Fig. \ref{fig:salto}. La
segunda derivada de $U_{ef}(x)$ en $x=0$, dada por
$U_{ef}''(x)=-6B(T_C-T)$, cambia de signo cuando $T=T_C$.

La aparici\'on de $U_{ef}$ se interpreta diciendo que la fuerza externa de
alta frecuencia hace oscilar el valor del potencial original $U$, pero el
valor medio de dicha oscilaci\'on puede cambiar la estabilidad de la
posici\'on de equilibrio de $U$ \cite{dittrich}.

Cuando $T$, o sea $y_0$, es grande, la masa $m$ oscila alrededor de la
posici\'on de equilibrio $x_0$. En este caso el movimiento mantiene la
simetr\'{\i}a con relaci\'on a la sustituci\'on $x\to -x$. Cuando $T$
baja hasta un valor $T<T_C$, la esfera oscilar\'a alrededor de una de las
dos posiciones de equilibrio $x_\pm$, en este caso el movimiento pierde
la mencionada simetr\'{\i}a.

El comportamiento de las frecuencias de oscilaci\'on como funciones de $T$
es matem\'aticamente an\'alogo a las transiciones de fase de segunda clase
que se estudian en f\'{\i}sica estad\'{\i}stica \cite{landauest}. Todos
los sistemas que presentan transiciones de fase poseen coordenadas
macrosc\'opicas, como la resistividad, el calor espec\'{\i}fico o la
magnetizaci\'on de una substancia, las cuales son an\'alogas de la
coordenada lenta $x$ de este problema. Las oscilaciones forzadas r\'apidas
constituyen el an\'alogo del fen\'omeno de transporte de calor y $T=y_0^2$
es an\'alogo de la temperatura. $T_C$ es el an\'alogo de la temperatura en
la cual ocurre la transici\'on de fase. En los alrededores de $T=T_C$ la
magnitud co\-rres\-pon\-dien\-te a $x_\pm$ (la resistividad por ejemplo)
es peque\~na, $x_\pm\propto(T_C-T)^{1/2}$ as\'{\i} como tambi\'en lo es la
frecuencia de las oscilaciones propias. En rigor, en el modelo mec\'anico
considerado se tiene una {\it bifurcaci\'on\/} que se comporta de manera
an\'aloga a una {\it transici\'on de fase\/}.

Este sistema pertenece a la clase de universalidad de los sistemas que
presentan transiciones de fase de segunda clase, seg\'un la denominaci\'on
de Landau. All\'{\i} se define $\epsilon=(T-T_C)/T_C$, el cual en este
caso es una funci\'on de $y_0$, $l$ y $d$. $x_\pm$ es el ``par\'ametro de
orden'' que depende de $\epsilon$ como
\begin{equation}
x_\pm=\pm\left(\frac{3T_C}{2}\right)^{1/2} \epsilon^{1/2},
\end{equation}
con esto el exponente del par\'ametro de orden es $\beta=1/2$. Por otra
parte la fuerza en $T=T_C$ vale $F_{ef}(T_C)=-4 B x^3$, con lo cual el
``exponente cr\'{\i}tico est\'a dado por $\delta=3$. Igualmente se pueden
encontrar otros par\'ametros an\'alogos a los que aparecen en transiciones
de fase.

En esta clase de transiciones ocurre un cambio de simetr\'{\i}a al variar
de manera continua un par\'ametro. Usualmente la ``fase'' que se presenta
para $T>T_C$ tiene mayor simetr\'{\i}a que la de $T<T_C$. Entre los
fen\'omenos de tran\-si\-ci\'on de fase de segunda clase est\'an: La
ferroelectricidad de las perovskitas \cite{landauest}, el ferromagnetismo
\cite{landauest}, la superconductividad \cite{bcs}, las oscilaciones en
plasmas \cite{anderson}, y el mecanismo por el cual los bosones en las
teor\'{\i}as de campos de norma de Yang-Mills adquieren masa \cite{anderson},
entre otros. Tambi\'en las oscilaciones transversas de una part\'{\i}cula
cargada en un campo magn\'etico no homog\'eneo constituyen un modelo
mec\'anico de estas transiciones, v\'ease el problema 10.8 en la
referencia \cite{serbo}.

\section{\textcolor{red}{El principio de invariancia de norma}}
\label{sec:gauge}

Queremos ahora hacer compatible el comportamiento descrito con las
pro\-pie\-da\-des del campo electromagn\'etico bajo transformaciones
de norma.

\subsection*{Ecuaciones de Maxwell}
El campo electromagn\'etico est\'a descrito por las ecuaciones de Maxwell,
las cuales en el sistema internacional de unidades toman la forma,

\begin{equation}
\begin{array}{rclrcl}
\vec{\nabla}\cdot\vec{D}&=&\rho,&\vec{\nabla}\cdot\vec{B}&=&0,\\
&\\
\vec{\nabla}\times\vec{E}&=&\displaystyle-\frac{\partial\vec{B}}{\partial
t},&
\vec{\nabla}\times\vec{H}&=&\displaystyle\vec{j}+\frac{\partial\vec{D}}{\partial
t};\\
&\\
\vec{D}&=&\epsilon_0\vec{E},&\vec{B}&=&\mu_0\vec{H}.
\end{array}
\end{equation}
\subsection*{Potenciales electromagn\'eticos}

\begin{equation}
\vec{B} = \vec{\nabla}\cdot\vec{A},\
\vec{E} = -\vec{\nabla}\phi -\frac{\partial\vec{A}}{\partial t}.
\end{equation}

\subsection*{Transformaciones de norma}
\begin{equation}
\vec{A}' = \vec{A}-\vec{\nabla}\chi,\
\phi' = \phi+\frac{\partial\chi}{\partial t}.
\end{equation}
$(\vec{A},\phi)$ y $(\vec{A}',\phi')$ dan lugar a los mismos campos
$(\vec{E},\vec{B})$.

\subsection*{Transformaciones can\'onicas}
Si se tiene un sistema mec\'anico cuyo hamiltoniano no depende del tiempo
y una transformaci\'on can\'onica $(q,p)\to(Q,P)$, se cumple
\begin{equation}
p_i=\frac{\partial F}{\partial q_i},\quad P_i=-\frac{\partial F}{\partial 
Q_i},\quad H'(Q,P)=H(q,p),
\label{eq:generatriz}
\end{equation}
donde $F(q,Q)$ es la funci\'on generatr\'{\i}z de la transformaci\'on.

El lagrangiano est\'a relacionado con el hamiltoniano por
\begin{equation}
L=\sum\limits_i p_i \dot{q}_i-H.
\end{equation}
En consecuencia, al realizar la transformaci\'on se obtiene un nuevo
la\-gran\-gia\-no que est\'a relacionado con el viejo as\'{\i},
\begin{equation}
L(q,\dot q)=L'(Q,\dot Q)+\frac{d F}{d t},
\label{eq:LagTrCa}
\end{equation}
lo cual dice que en una transformaci\'on can\'onica el lagrangiano se
afecta adicionando la derivada total respecto al tiempo de la funci\'on
generatr\'{\i}z de la transformaci\'on, que como se sabe no cambia la
din\'amica del problema. Una transformaci\'on puntual es tambi\'en
can\'onica.

\subsection*{Notaci\'on 4-dimensional}
Definimos $x_4=ict$ y el tensor m\'etrico $g_{\mu\nu}=\delta_{\mu\nu}$.
Entonces la corriente es $j_\mu=(\vec{j},ic\rho)$, la cual cumple la
ecuaci\'on de continuidad $\partial^\mu j_\mu=0$. $\rho$ es la densidad
de carga. El potencial vectorial es $A_\mu=(\vec{A},i\phi/c)$. El tensor
de campo electromagn\'etico es \cite{landaucam},
\begin{equation}
F_{\mu\nu} = 
\frac{\partial A_\nu}{\partial x^\mu}-
\frac{\partial A_\mu}{\partial x^\nu}.
\end{equation}

Las ecuaciones de Maxwell expresadas en t\'erminos de los campos toman la
forma
\begin{equation}
\begin{array}{rcl}
\partial^\nu F_{\mu\nu}&=&\mu_0 j_\mu,\\
&&\\
\partial_\mu F_{\nu\lambda}+
\partial_\nu F_{\lambda\mu}+
\partial_\lambda F_{\mu\nu}&=&0.
\end{array}
\end{equation}
Las ecuaciones de Maxwell expresadas en t\'erminos de los potenciales
toman la forma
\begin{equation}
\Box A_\nu -\partial_\nu\partial^\mu A_\mu
=\mu_0 j_\nu,
\label{eq:dalambert}
\end{equation}
y la otra ecuaci\'on se satisface autom\'aticamente por la expresi\'on de
los campos en t\'erminos de los potenciales.

Se pueden elegir convenientemente los potenciales. Usualmente se usa la
invariancia de norma para restringir la clase de potenciales a usarse, por
ejemplo la clase de los que satisfacen la condici\'on de Lorentz
\begin{equation}
\partial_\mu A^\mu=0.
\end{equation}
Un paso adicional, la elecci\'on de la norma de Coulomb, escoge
un\'{\i}vocamente el potencial dentro de los de dicha clase,
\begin{equation}
A_4=0,\ \nabla\cdot\vec{A}=0.
\end{equation}

\subsection*{La part\'{\i}cula cargada en un campo EM}
El lagrangiano (no relativista),
\begin{equation}
L=\frac{1}{2}m\vec{v}^2-q\phi+q\vec{v}\cdot\vec{A}.
\label{eq:lagEM}
\end{equation}

Ecuaci\'on de Lagrange,
\begin{equation}
\frac{d}{dt}(m\vec{v})=q(\vec{E}+\vec{v}\times\vec{B}).
\end{equation}

Momento can\'onico,
\begin{equation}
\vec{p}=m\vec{v}+q\vec{A}.
\end{equation}

Hamiltoniano,
\begin{equation}
H=\frac{1}{2m}(\vec{p}-q\vec{A})^2+q\phi.
\end{equation}

El lagrangiano en un sistema de referencia que rota con velocidad angular 
definida por $m\vec\omega\times\vec{r}=-q\vec{A}$, al aplicar la
transformaci\'on $\vec{v}=\vec{v}_r+\vec\omega\times\vec{r}$ es,
\begin{equation}
L_r=\frac{1}{2}m\vec{v}_r^2-\frac{q^2}{2m}\vec{A}^2-q\phi.
\end{equation}
Cuando $\vec{B}$ es homog\'eneo se tiene
$\vec{A}=-\vec{r}\times\vec{B}/2$ y por tanto
$\vec\omega=-q\vec{B}/(2m)$, la frecuencia de Larmor. En este caso,
\begin{equation}
L_r=\frac{1}{2}m\vec{v}_r^2-\frac{q^2}{8m}\vec{B}^2(x^2+y^2)-q\phi.
\label{eq:larmor}
\end{equation}

\subsection*{Acci\'on e invariancia bajo transformaciones de norma}
Si los potenciales que aparecen en el lagrangiano de la ecuaci\'on
(\ref{eq:lagEM}) se someten a una transformaci\'on de norma, este es
reemplazado por
\begin{equation}
\begin{array}{rcl}
L'&=&\displaystyle\frac{1}{2}m\vec{v}^2-q(\phi+\frac{\partial\chi}{\partial 
t})+q\vec{v}\cdot(\vec{A}-\nabla\chi)\\
&&\\
&=&\displaystyle\frac{1}{2}m\vec{v}^2-q\phi + q\vec{v}\cdot\vec{A}
-q\frac{d\chi}{d t}.
\end{array}
\end{equation}

La acci\'on de Hamilton es
\begin{equation}
S=\int\limits_0^t L dt.
\end{equation}

Si la funci\'on arbitraria $\chi$ se escoge tal que $\chi=0$ en $t=0$,
entonces la acci\'on de Hamilton se transforma en
\begin{equation}
S'=S-q\chi.
\end{equation}
Por propiedades de la acci\'on de Hamilton se tiene que la din\'amica de
la part\'{\i}cula es independiente de la elecci\'on de la norma.

Si la funci\'on de onda semicl\'asica es \cite{landaucua}
\begin{equation}
\Psi=a e^{iS/\hbar},
\end{equation}
se obtiene que un cambio de norma produce un cambio en la fase, as\'{\i},
\begin{equation}
\Psi'=\Psi e^{-iq\chi/\hbar}.
\end{equation}

Una transformaci\'on de norma en los potenciales electromagn\'eticos no
tiene consecuencias observables en mec\'anica cl\'asica porque s\'olo da
lugar a la adici\'on a la lagrangiana de la derivada total respecto al
tiempo de una funci\'on arbitraria de las coordenadas espacio-temporales.
Esto est\'a asociado a un cambio de fase de la funci\'on de onda, que en
mec\'anica cu\'antica tampoco tiene efectos observables. El argumento
puede invertirse: La adici\'on de la derivada total de una funci\'on
arbitraria implica la presencia de un campo vectorial $A_\mu$ que
experimenta una transformaci\'on de norma. Esta es la forma cl\'asica del
principio de invariancia de norma. En mec\'anica cu\'antica se tiene que el tipo de
interacci\'on est\'a determinado por los cambios de fase de la funci\'on
de onda.

Si se realiza una transformaci\'on can\'onica de las variables
mec\'anicas del sistema, de acuerdo a la ecuaci\'on (\ref{eq:LagTrCa}) se
adiciona la derivada total respecto al tiempo de la funci\'on
generatr\'{\i}z. En consecuencia, al combinar los dos efectos se puede
decir que el lagrangiano quedar\'a invariante (e igualmente la fase de la
funci\'on de onda semicl\'asica) si se realiza una transformaci\'on de
norma tal que
\begin{equation}
q\chi=F.
\label{eq:GaugeTC}
\end{equation}
Como $F$ es funci\'on de las coordenadas, se tiene que cada
transformaci\'on can\'onica puede compensarse con una transformaci\'on
de norma {\it local\/}. Si esto es as\'{\i}, el lagrangiano queda invariante
y se tiene que esa combinaci\'on de transformaciones representa una {\it
simetr\'{\i}a del lagrangiano\/}.

En conclusi\'on, exigir invariancia de norma bajo transformaciones
can\'onicas infinitesimales (las cuales son locales en el espacio de
fases) requiere la introducci\'on de una interacci\'on con el campo
electromagn\'etico. La invariancia de norma local genera din\'amica.

\section{\textcolor{red}{Simetr\'{\i}a oculta}}
\label{sec:hidden}

Sea el problema de la part\'{\i}cula cargada que se mueve con la
energ\'{\i}a potencial $U(\vec{\rho})=-C\rho^2+B\rho^4$. Puede
considerarse como una generalizaci\'on del sistema de la figura
\ref{fig:resortes}. Suponemos que este $U$ no es de naturaleza
electromagn\'etica. En nuestro an\'alisis tampoco son de naturaleza
electromagn\'etica las oscilaciones de frecuencia r\'apida que determinan
el tipo de estabilidad del punto con $\rho=0$. $U(\vec{\rho})$ es una
superficie con la forma de un sombrero mexicano. S\'olo se consideran
movimientos en el plano $x-y$. $U(\rho)$ no depende del \'angulo $\phi_e$
que hace $\rho$ con el eje $x$, $H$ tiene la simetr\'{\i}a de rotaci\'on
alredededor del eje $z$ por un \'angulo arbitrario.

Como se vi\'o, el estado de equilibrio estable se alcanza no en $\rho=0$,
donde $U=0$, sino sobre un c\'{\i}rculo de radio $\rho_e=\sqrt{C/(2B)}$,
donde $U$ es negativo. Es claro que cualquier punto del c\'{\i}rculo es
una posible posici\'on de equilibrio. Todos esos puntos son igualmente
estables y aceptables como ``el'' estado base, difieren s\'olo por la
``fase'' $\phi_e$. Esos ``estados base'' que difieren s\'olo por $\phi_e$
son equivalentes, ah\'{\i} reside la simetr\'{\i}a. El ``rompimiento
espont\'aneo de la simetr\'{\i}a'' ocurre cuando se escoge un valor
particular de $\phi_e$. As\'{\i}, si se toma $\phi_e=0$ se tiene
$\vec\rho_e=\sqrt{C/(2B)}\hat{\imath}$.

Es evidente de la ecuaci\'on (\ref{eq:GaugeTC}) que al tomar el
equilibrio en
\begin{equation}
\vec\rho_e=\left(\frac{C}{2B}\right)^{1/2}(\hat{\imath}\cos\phi_e +
\hat{\jmath}\sin\phi_e),
\end{equation}
con $\phi_e$ arbitrario se produce un cambio en el lagrangiano que puede
compensarse con una transformaci\'on de norma sobre $A_\mu$, y por lo tanto
dicha elecci\'on es una simetr\'{\i}a del lagrangiano.

Por efecto de un campo magn\'etico $\vec B$ se induce un momento
magn\'etico $\vec\mu$ que a su vez interact\'ua con el campo para dar la
energ\'{\i}a adicional $E=-\vec\mu\cdot\vec{B}$, cuyo valor es
\begin{equation}
\frac{q^2}{8m}\vec{B}^2(x^2+y^2),
\end{equation}
que coincide con el t\'ermino dependiente de $B$ en la ecuaci\'on
(\ref{eq:larmor}).

Como consecuencia del diamagnetismo, el campo magn\'etico que
ex\-pe\-ri\-men\-ta la carga no coincide con $\vec{B}$, sino que es menor
debido al campo de direcci\'on opuesta que se crea de acuerdo a la ley de
Lenz, ``el campo diamagn\'etico interno''. El momento magn\'etico es
proporcional a la corriente y esta a la densidad de carga.

El diamagnetismo est\'a asociado a las llamadas corrientes de
a\-pan\-ta\-lla\-mien\-to diamagn\'etico, las cuales a su vez dan lugar a
la aparici\'on de un ``t\'ermino de masa'' en la ecuaci\'on de onda para
el fot\'on. $\vec{j}_{scr}$ es proporcional al potencial vectorial,
\begin{equation}
\vec{j}_{scr}=-\frac{q\rho}{m}(\vec{A}-\vec\nabla\phi),
\label{eq:screen}
\end{equation}
donde $\rho$ es la densidad de carga, en este caso dada por una $\delta$
de Dirac por tenerse una part\'{\i}cula puntual. Puede generalizarse par
incluir objetos no puntuales como cuerdas o membranas. Esta ecuaci\'on
(sin el t\'ermino de la norma) es an\'aloga a la de London de la
superconductividad.

La ecuaci\'on de Maxwell (\ref{eq:dalambert}) para $A_\nu$, luego de
imponer la condici\'on de Lorentz, y de utilizar una norma dependiente de
$\phi_e=-\tan^{-1}(y/x)$, toma la forma \begin{equation} (\Box + M^2)A_\nu
= 0, \label{eq:masa} \end{equation} donde $M^2=q\rho/m$. Es la ecuaci\'on
de Proca. Comp\'arese con el resultado de \cite{ogawa}. La ecuaci\'on
(\ref{eq:masa}) representa el campo de una part\'{\i}cula libre vectorial
de masa $M$. El sistema original tiene cuatro grados de libertad, dos
modos de oscilaci\'on de la part\'{\i}cula y dos estados de polarizaci\'on
del campo. El rompimiento espont\'aneo de simetr\'{\i}a y la elecci\'on
apropiada de la norma han dado lugar a un sistema equivalente en el cual la
part\'{\i}cula tiene s\'olo un modo de oscilaci\'on pero el campo tiene
tres estados de polarizaci\'on. A este resultado se llega f\'acilmente, en
estrecha analog\'{\i}a con el campo bos\'onico cargado interactuante con
el campo EM, a partir de la densidad lagrangiana del sistema formado por
el campo y la part\'{\i}cula \cite{halzen}. Tambi\'en es relevante
para el presente modelo
una discusi\'on an\'aloga a la de \cite{aitchison} acerca del t\'ermino de masa.

\section{\textcolor{red}{Conclusi\'on}}
\label{sec:concl}
Se analiz\'o un sistema constituido por una part\'{\i}cula en un potencial
$U(\rho)$. Se encontr\'o que se presentan tres fen\'omenos relacionados:
La existencia de una ``transici\'on de fase'', m\'as precisamente de una
bifurcaci\'on, que se puede atribuir a un campo r\'apidamente oscilante,
en la cual el origen de coordenadas cambia su tipo de estabilidad. A su
vez, cuando el origen se desestabiliza aparece un conjunto de nuevas
posiciones estables equivalentes relacionadas entre si por el grupo
$U(1)$. Una posici\'on de equilibrio cualquiera se puede mapear en la
localizada sobre el eje $x$ al escoger la norma del potencial vectorial
correspondiente al \'angulo de la rotaci\'on requerida (``rompimiento de
la simetr\'{\i}a''). La escogencia de esta norma da lugar a la aparici\'on
de un ``t\'ermino de masa'' en la ecuaci\'on de onda que determina el
potencial vectorial. Esta masa depende de las corrientes diamagn\'eticas
del sistema. A su vez las oscilaciones de la part\'{\i}cula cargada
alrededor de la nueva posici\'on de equilibrio representan el ``campo de
Higgs'' y su masa es positiva. V\'ease el comentario que sigue a la
ecuaci\'on (\ref{eq:erroneo}).

Una conclusi\'on de este ejercicio es que el sistema analizado podr\'{\i}a
tomarse como un ``sistema modelo'' del principio de invariancia de norma y sus efectos
conexos. Una realizaci\'on macrosc\'opica del modelo puede ser
dif\'{\i}cil a causa de la peque\~nez de las corrientes diamagn\'eticas.
Fen\'omenos de este tipo se podr\'{\i}an presentar en estados de Rydberg
de un \'atomo hidrogenoide en un campo magn\'etico. Tambi\'en campos de norma
no abelianos pueden encontrarse en sistemas de espines cl\'asicos en
campos magn\'eticos.


\begin{thebibliography}{99}
\bibitem{acus} A. Acus, ``Can spontaneous symmetry breaking occurs in a
potential with one minimum?'' physics/0003030.
\end{thebibliography}
\end{document}